\documentstyle[sprocl,epsfig,bm]{article}

\def\NPB{{\em Nucl. Phys.} B}

\def\PRD{{\em Phys. Rev.} D}

\begin{document}

\title{THE $\bm{e^- e^-}$ MODE OF LC: OPPORTUNITIES TO DISCOVER LOOP-LEVEL LEPTON FLAVOR (NUMBER) VIOLATION}

\author{MIRCO CANNONI }

\address{LPNHE, Laboratoire de Physique Nucl\'eaire et de Hautes Energies,\\
IN2P3 - CNRS, Universit\'e Paris VI et VII, 4 Place Jussieu, 75525 Paris cedex 05, France}

\author{ORLANDO PANELLA AND STEPHAN KOLB}

\address{INFN, Sezione di Perugia, Via G. Pascoli 1, 06129, Perugia, Italy}

\maketitle
\abstracts{
The $e^- e^-$ mode with lepton number $L=+2$ of the initial state is particularly suitable 
for studying lepton and flavor number violating reactions (LFV). 
In this brief report we give a summary of the study of two loop-level reactions 
which violate lepton and flavor numbers:
we first consider seesaw type models with heavy Majorana neutrinos at the TeV scale
and study the reactions $e^{-}e^{-} \to \ell^{-}\ell^{-}$,
and then similar reactions $e^{-}e^{-} \to \ell^{-}e^{-}$ ($\ell=\mu,\tau$) in supersymmetric
models where LFV is due to slepton mixing. 
More details on the calculations and numerical
tools used can be found in Refs.~\cite{Cannoni2002} while summary of the theoretical scenarios
is presented in the summary report of the $e^- e^-$ session~\cite{Cheung}.
}

\section{$\bm{e^{-}e^{-}\;\to\;\ell^{-}\ell^{-}}$ ($\bm{\ell=\mu,\tau}$) With TeV Scale Majorana Neutrinos  }
To produce a detectable signal, heavy Majorana neutrinos (HMN), besides having masses in the TeV range, 
must have
interactions which are not suppressed by the mixing matrices as it happens in the one 
family seesaw mechanism, where $\theta\ {\simeq}\ \sqrt{{m_{\nu}}/{M_N}}$. 
With three generations, more free parameters are at our disposal, 
and the ``two miracles'' of not so large masses 
and non negligible mixing, are obtained imposing suitable relations among the elements of the matrices $m_D$
and $M_R$: examples of these models are proposed in Refs.~\cite{buchmuller91}. 
Experimentally one cannot put bounds on single mixing matrix elements, but on some combinations 
of them, assuming that each charged lepton couples only to one heavy neutrino with significant strength.
Light-heavy mixing has to be inferred
from low-energy phenomenology
and from global fits performed
on LEP data identifying the following effective mixing angles 
$s^{2}_{\ell} =\sum_{j} |B_{\ell_i N_j}|^{2}{\equiv}\sin^{2}{\theta}_{{\nu}_{\ell}}$
with upper bounds~\cite{Nardi}:
$s^{2}_{{e}}< 0.0054$,
$s^{2}_{{\mu}}< 0.005$,
$s^{2}_{{\tau}}< 0.016$,
where $B$ is the mixing matrix appearing in the charged current weak interaction lagrangian.
Under these assumptions, the coupling of neutrissimos to gauge bosons and leptons is numerically fixed 
to $gB_{\ell N_{i}}$.
Since the width of the heavy states grows as $M^{3}_N$, at a certain value it will 
happen that $\Gamma_N > M_N$, signaling a 
breakdown of 
perturbation theory. 
The perturbative limit on $M_N$ is thereby estimated requiring 
$\Gamma_N < M_N/2$, which gives an upper bound of 
$\simeq 3$ TeV for the numerical values of the mixing used.
We study the process $e^{-}e^{-}\to \ell^{-}\ell^{-}$ ($\ell=\mu,\tau$),
described by box diagrams with two HMN and two $W$ gauge bosons in the loop.
\begin{figure}[t!] 
\begin{center}
\epsfig{file=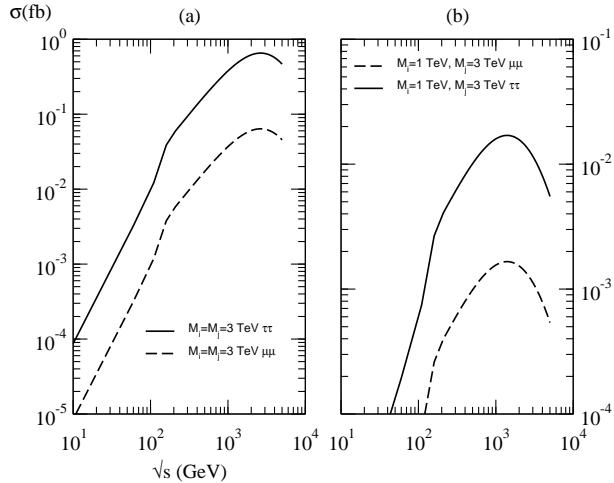,width=8cm}
\end{center}
\caption{Total unpolarized cross sections as function of $\sqrt{s}$ for some values of HMN masses.}  
\label{fig1}
\end{figure}
One expects an enhancement of the cross section at $\sqrt{s}\simeq 161$ GeV $\simeq{2M_{W}}$,
the threshold for on-shell $W W$ gauge boson production, at which 
the four-point functions develop an imaginary part.
The enhancement due to the threshold singularity of the loop amplitude
is more pronounced for values of Majorana masses close to $M_W$ and 
is drastically reduced increasing $M_{N_i} \approx M_{N_j}$ to  
${\cal O}$(TeV). 
The main contribution comes 
from the graph with two Goldstone bosons (the calculation is performed in the 't Hooft-Feynman gauge)
since their coupling is proportional to $M_{{N}_{i}}$. Moreover the chiral 
structure of the coupling selects the 
mass term in the numerator of the Majorana neutrino propagators. 
When these masses are much larger then the other quantities, 
the amplitude scales like $M^{3}_{{N}_{i}} 
M^{3}_{{N}_{j}}/M^{2}_{{N}_{i}}M^{2}_{{N}_{j}}\simeq M_{{N}_{i}}M_{{N}_{j}}$, 
i.e. is proportional to the square of the heavy masses.
As shown in Fig.~\ref{fig1}, (where both the two previous effects are easily seen)
for $M_{N_i} = M_{N_j} = 3 $ TeV the signal does reach the 
level of $10^{-1},10^{-2}$ fb respectively 
for the ($\tau\tau$) and the ($\mu\mu$) signals at $\sqrt{s}= 500 $ GeV, 
which for an annual integrated luminosity of $100$ fb$^{-1}$ would 
correspond respectively to 10 and 1 event/year. At higher energies,
$\cal O$ (TeV), one could get even larger event rates (30 and 3) respectively: 
this is largest because the upper limits on the mixing are less stringent.
In the limit of massless external particles the process is dominated by a well defined 
helicity amplitude: $ e_L e_L \to \ell_L \ell_L$. With left polarized initial beams
one can enhance the signal by a factor of four (for ideal 100\% polarization). 

\section{$\bm{e^{-}e^{-}\;\to\; \ell^{-}e^{-}}$ ($\bm{\ell=\mu,\tau}$) 
In R-Conserving Supersymmetric Models }

The diagonalization of slepton mass matrices induce LFV couplings in the 
lepton-slepton-gaugino vertices: off diagonal entries in slepton mass matrices are 
generated, for example, by the seesaw mechanism embedded in the MSSM with R-parity conservation
and mSUGRA boundary conditions at high energy.  
Anyway our approach
is model independent: in order to keep the discussion simple enough, 
the maximal mixing of only two generations is considered, and accept the 
flavor violating entries of slepton mass matrices as large as allowed by the experimental 
bounds on rare LFV deacays.
The essential parameter which controls the LFV signal is
$\delta_{LL}=\frac{\Delta m^{2}}{\tilde{m}^{2}}$.
It is assumed that
the two lightest neutralinos are pure bino and pure wino 
with masses $M_1$ and $M_2$ respectively, while 
charginos are pure charged winos with mass $M_2$, $M_1$ and $M_2$  
being the gaugino masses in the soft breaking potential.
Numerical results are obtained using the mSUGRA relation 
$M_1 \simeq 0.5 M_2$ for gaugino masses while $\Delta m^{2}$ and the slepton 
masses are taken to be free phenomenological parameters. 
The are three contributing amplitudes:
${\cal M}_{E1}={\cal M}(e^-_L e^-_L \to \ell^-_L e^-_L)$,
${\cal M}_{E2}={\cal M}(e^-_L e^-_R \to \ell^-_L e^-_R)$,
${\cal M}_{E3}={\cal M}(e^-_R e^-_L \to \ell^-_L e^-_R)$.
${\cal M}_{E1}$ has $J_z=0$, is flat and forward-backward 
symmetric because of the antisymmetrization.
${\cal M}_{E2}$ and ${\cal M}_{E3}$ describe
P-wave scattering with $J_z=+1$ and $J_z=-1$ respectively and 
are peaked in the forward direction and in the backward direction.
The signal cross section is dominated 
by the amplitude ${\cal M}_{E1}$. 
The analysis of the corresponding total 
cross section as a function of $\sqrt{s}$ 
shows that at $\sqrt{s}=2\tilde{m}_L$ 
$\sigma$ changes of orders of magnitude 
giving a sharp peak that is smeared only by large values of 
$\Delta m^2$.
This can be easily understood considering the threshold behavior of the 
cross section for slepton pair production~\cite{Feng2}: 
defining $\beta=\sqrt{1-4 m^{2}_{\tilde{L}}/s}$ the selectron velocity,
the amplitude of the intermediate state $e^-_L e^-_L \to \tilde{e}^-_L \tilde{e}^-_L$ behaves like 
$\beta$, while for the other two cases it goes like $\beta^3$.
With SUSY masses not much larger than $\sim 200$ GeV the signal is of order 
${\cal O}$(10$^{-2}$) fb for $\delta_{LL} > {\cal O}(10^{-1})$. 
In addition the cross section is practically angle independent and thus 
insensitive to angular (or tranverse momentum) cuts.
The phenomenological points of the SUSY parameter space corresponding to gaugino 
masses $(M_1,\;M_2)=(80,\;160)$ GeV or  $(100,\;200)$ GeV and to 
slepton masses $m_L=100-200$ GeV and $\delta_{LL}>10^{-1}$
(which implies $\Delta m^2>10^{3}$ GeV$^{2}$) can give in the 
$e^- e^-$ mode a detectable LFV signal ($e^-e^- \to \ell^- e^-$)  
although at the level of ${\cal O}(1-25)$ events/yr with 
${L}_0=100$ fb$^{-1}$. Higher sensitivity to the SUSY 
parameter space could be obtained with larger $L_0$. 
\begin{figure}[t!] 
\begin{center}
\epsfig{file=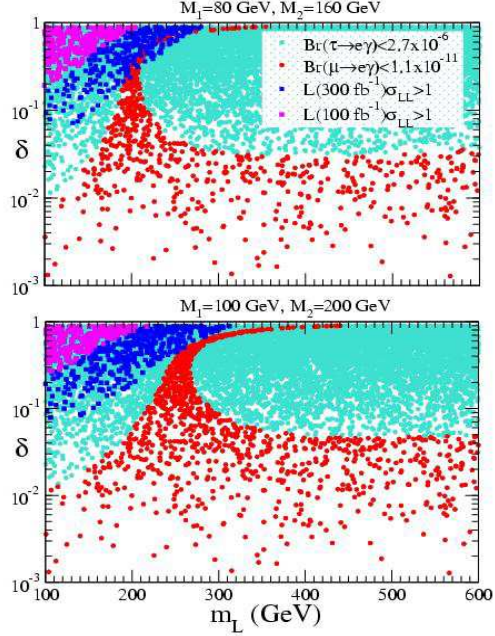,width=6.5cm,height=8.5cm}
\end{center}
\caption{Scatter plot in the plane ($\delta_{LL}, m_L$) of: (a) the 
experimental bounds from $\mu\to e\gamma$ and 
$\tau \to \mu \gamma$ (allowed regions with circular dots); (b) regions where the signal 
can give at least one event with two different values of integrated 
luminosity (squared dots), for two sets of gaugino masses. Each signal point is calculated 
at $\sqrt{s}=2\tilde{m}_L$.} 
\label{fig2}
\end{figure} 
On the other hand the experimental bounds on rare lepton 
decays $\mu ,\tau \to e\gamma$ set constraints on the LFV violating 
paramters $\Delta \tilde{m}^2$ or $\delta_{LL}$.
Fig.~\ref{fig2} shows that the bound from $\tau \to e \gamma$ does not 
constrain the region of the ($\delta_{LL}, m_L$) plane
compatible with an observable LFV signal and therefore
the reaction $e^-e^- \to \tau^- e^-$ could produce a detectable signal
whithin the highlighted regions of the parameter 
space. 
The process $e^-e^- \to \mu^- e^-$ is 
observable only in a small section of the parameter space
since the allowed region from the $\mu \to e \gamma$ decay 
almost does not overlap with the collider ``discovery'' region except 
for a very small fraction in the case of gaugino masses ($M_1=80$ GeV and 
$M_2=160 $ GeV). The compatibility of values of 
$\delta_{LL} \approx 1$ is due to a cancellation among the  
diagrams that describe the $\ell \to \ell' \gamma$ decay in particular points 
of the parameter space.

\section{Conclusions}

In summary, using the maximum experimentally allowed mixings, 
that masses of heavy Majorana neutrinos 
up to $2-3$ TeV can be explored with the reaction $e^- e^- \to  \ell^- \ell^-, (\ell = \mu, \tau)$, because 
the amplitude gets an enhancement at the threshold for two gauge bosons production
and then shows a non-decoupling behavior with the mass of the virtual heavy states.
For the similar reaction $e^- e^- \to  \ell^- e^-, (\ell = \mu, \tau)$ induced by slepton mixing in 
supersymmetric models,
in certain regions of the parameter space, the signal can reach the level of $10^{-2}$ fb around the 
threshold for selectrons pair production. 
The possibility of employing beams with high degree of 
left longitudinal polarization is also essential to enhance the signal. 
These signals have the unique characteristic of a back to back high 
energy lepton pair and no missing energy. 
Sources of Standard Model background like  
the reaction $e^{-}e^{-} \to \nu_{e}\nu_{e} {W^{-}}^{*} {{W}^{-}}^{*}$
followed by the decays ${W^{-}} {{W}^{-}} \to 
\ell^{-} {\bar{\nu}}_{\ell}{\ell^{-}}' {\bar{\nu}}_{{\ell}'}$,
are studied in details in Ref.~\cite{Cannoni2002} where it is shown
that with reasonable cuts on the transverse momenta of the leptons 
and on the missing energy, it will be drastically reduced, 
without affecting significantly the signal.

M.~C. thanks: ``Fondazione Angelo Della Riccia'' for a fellowship, the Theory group of Department of Physics of 
University of Perugia for partial support; C.~Carimalo and LPNHE in Paris for the kind hospitality.
The work of S.~K.~at INFN-Perugia in the period 2001-2002 has been supported by 
European Union under the Contract No.~HPMF-CT-2000-00752.

\end{document}